\date{October 2020}

\documentclass[twocolumn,aps,prx,longbibliography,amsmath,amssymb,superscriptaddress]{revtex4-2}

\usepackage{graphicx}
\usepackage{dcolumn}
\usepackage{bm}
\usepackage{float}
\usepackage{hyphenat}
\usepackage{hyperref}
\usepackage[dvipsnames]{xcolor}
\usepackage{physics}
\usepackage[T1]{fontenc}

\newcommand{\heriotwatt}{Institute of Photonics and Quantum Sciences, SUPA, Heriot-Watt University, Edinburgh EH14 4AS, UK}
\newcommand{\TsukubaKenji}{Research Center for Functional Materials, National Institute for Materials Science, 1-1 Namiki, Tsukuba 305-0044, Japan}
\newcommand{\TsukubaTakashi}{International Center for Materials Nanoarchitectonics, National Institute for Materials Science,  1-1 Namiki, Tsukuba 305-0044, Japan}

\begin{document}

    \title{Moir{\'e}-trapped interlayer trions in a charge-tunable WSe$_2$/MoSe$_2$ heterobilayer}
    
    \author{Mauro Brotons-Gisbert}
    \email{M.Brotons_i_Gisbert@hw.ac.uk}
    \affiliation{\heriotwatt}
    \author{Hyeonjun Baek}
    \affiliation{\heriotwatt}
    \author{Aidan Campbell}
    \affiliation{\heriotwatt}
    \author{Kenji Watanabe}
    \affiliation{\TsukubaKenji}
    \author{Takashi Taniguchi}
    \affiliation{\TsukubaTakashi}
    \author{Brian D. Gerardot}
    \email{B.D.Gerardot@hw.ac.uk}
    \affiliation{\heriotwatt}
    
    \date{\today}

    \begin{abstract}
    
        Transition metal dichalcogenide heterobilayers offer attractive opportunities to realize lattices of interacting bosons with several degrees of freedom. Such heterobilayers can feature moir{\'e} patterns that modulate their electronic band structure, leading to spatial confinement of single interlayer excitons (IXs) that act as quantum emitters with $C_3$ symmetry. However, the narrow emission linewidths of the quantum emitters contrast with a broad ensemble IX emission observed in nominally identical heterobilayers, opening a debate regarding the origin of IX emission. Here we report the continuous evolution from a few trapped IXs to an ensemble of IXs with both triplet and singlet spin configurations in a gate-tunable $2H$-MoSe$_2$/WSe$_2$ heterobilayer. We observe signatures of dipolar interactions in the IX ensemble regime which, when combined with magneto-optical spectroscopy, reveal that the narrow quantum-dot-like and broad ensemble emission originate from IXs trapped in moir{\'e} potentials with the same atomic registry. Finally, electron doping leads to the formation of three different species of localised negative trions with contrasting spin-valley configurations, among which we observe both intervalley and intravalley IX trions with spin-triplet optical transitions. Our results identify the origin of IX emission in MoSe$_2$/WSe$_2$ heterobilayers and highlight the important role of exciton-exciton interactions and Fermi-level control in these highly tunable quantum materials.
    
    \end{abstract}

    \maketitle

    \section{Introduction}
    
    Van der Waals heterobilayers consisting of vertically stacked monolayer transition-metal dichalcogenide semiconductors (TMDs) form atomically sharp interfaces with type-II band alignment \cite{chiu2015determination,wilson2017determination}. This band alignment enables the formation of interlayer excitons (IXs) - electron-hole Coulomb bound states between electrons and holes spatially separated in different monolayers. The reduced overlap of the electron and hole wavefunctions gives rise to long IX radiative lifetimes (compared to intralayer excitons)  \cite{rivera2015observation, miller2017long} that can be further tailored by the momentum mismatch between the carriers \cite{choi2021twist}. The spatial separation of the IXs carriers also results in a large permanent electric out-of-plane dipole moment, which enables a large tunability of the exciton energy by externally applied electric fields \cite{ciarrocchi2019polarization,jauregui2019electrical,baek2020highly}. The combination of long lifetimes and large binding energies \cite{rivera2018interlayer,Torun2018} position IXs in TMD heterostructures as an exciting platform to explore many-body exciton-exciton phenomena such as dipolar interactions of IXs in the low density regime \cite{kremser2020discrete,li2020dipolar} or the high-density regime, where signatures of coherent excitonic many-body quantum states and high-temperature exciton (boson) condensation have been predicted \cite{fogler2014high,wu2015theory,berman2016high} and observed \cite{sigl2020signatures,wang2019evidence}. The robust and long-lived IXs in TMD heterobilayers also offer novel opportunities to realize atomically-thin optoelectronic devices such as lasers \cite{paik2019interlayer} and excitonic transistors that can operate at room temperature \cite{unuchek2018room}.
    
    Beyond the large permanent dipole moment and strong Coulomb interactions, the compelling concept of a moir{\'e} superlattice \cite{kang2013electronic} emerges in TMD heterobilayers due to the lattice mismatch and any relative twist angle between the constituent monolayers. In MoSe$_2$/WSe$_2$ heterobilayers, the moir{\'e} superlattice results in a periodic potential landscape for IXs \cite{zhang2017interlayer} (with a periodicity that depends on the relative crystallographic alignment of the layers) in which three trapping sites with three different local atomic registries arise \cite{yu2017moire,wu2018theory,yu2018brightened}. Experimental evidence of IXs trapped in a moir{\'e}-induced potential has been reported in MoSe$_2$/WSe$_2$ heterobilayers with twist angles near 0$^\circ$, 21.8$^\circ$ and 60$^\circ$ at cryogenic temperatures \cite{seyler2019signatures,brotons2020spin,baek2020highly}. These localized IXs present emission linewidths below 100 $\mu$eV \cite{seyler2019signatures,brotons2020spin,baek2020highly} and photon antibunching \cite{baek2020highly}, clear hallmarks of quantum-confined excitons. Moreover, the trapped IXs show well-defined magneto-optical properties: the $g$-factors of the trapped IXs depend on the relative valley alignment (i.e., stacking configuration) between the layers hosting the carriers, while their optical selection rules are determined by the atomic registry of the trapping site \cite{seyler2019signatures,brotons2020spin}. Together, these magneto-optical properties provide compelling evidence for the moir{\'e} potential as the origin of the IX trapping. The important role the moir{\'e} potential plays in the confinement of IXs is further supported by the twist-dependent IX diffusion in TMD heterobilayers, in which the diffusion length of the IX ensemble depends on the moir{\'e} periodicity \cite{choi2020moire,yuan2020twist}. However, the narrow emission linewidths observed for single trapped IXs contrast with the broad photoluminescence (PL) spectra observed in similar MoSe$_2$/WSe$_2$ heterostructures  \cite{hanbicki2018double,ciarrocchi2019polarization,calman2020indirect,wang2019giant,jauregui2019electrical,miller2017long,sigl2020signatures,choi2021twist}, which show IX emission bands with linewidths of $4-6$ meV in the cleanest samples \cite{ciarrocchi2019polarization,calman2020indirect,sigl2020signatures,delhomme2020flipping}, two orders of magnitude broader than single trapped IXs. Such contrasting observations have opened a debate regarding a unified picture of the nature of IX emission in TMD heterobilayers, and in particular in the prototypical heterostructure: MoSe$_2$/WSe$_2$ heterobilayers \cite{tartakovskii2020moire}. To date, clear experimental evidence which can marry these two regimes (narrow linewidth trapped IX vs broad linewidth ensemble IX) is missing. Further, gate doping of MoSe$_2$/WSe$_2$ heterobilayers has been shown to lead to the formation of charged IX (trions) with spin-singlet and spin-triplet configurations \cite{jauregui2019electrical,joe2021electrically}. However, the trapped or delocalized nature of IX trions, and all their possible spin-valley configurations, have yet to be addressed.
    
    Here, we investigate the magneto-optical properties of IXs in a gate-tunable MoSe$_2$/WSe$_2$ heterobilayer. We tune the density of IXs by scanning the excitation power over five orders of magnitude and report a clear and continuous evolution from the narrow emission of single trapped IXs to broad ensemble IX peaks, for which IXs with both spin-triplet and spin-singlet configuration are observed. In the high excitation power regime, we observe an energetic blue-shift of the IXs due to repulsive dipolar interactions. We estimate the density of IXs from the power dependent evolution of the IX PL and find that, even at the highest excitation powers employed in our measurements ($\sim$80 $\mu$W), the estimated density of optically-generated IXs is two orders of magnitude smaller than the estimated density of moir{\'e} traps ($\sim4\cdot10^{12}$ cm$^{-2}$) in our heterostructure. Polarization-resolved PL measurements under a vertical magnetic field confirm that the narrow and broad IX PL present identical magneto-optical properties, confirming that both the quantum-dot-like and broad PL emission from IXs arise from IXs trapped in moir{\'e} potentials with the same atomic registry. Moreover, we investigate the formation of negatively-charged IX trions and find that the trion creation originates from on-site charging of the trapping potentials. The magneto-optical properties of the negatively-charged IXs reveal three different negative trion species with contrasting spin-valley configurations. Interestingly, we observe both intervalley and intravalley IX trions with spin-triplet optical transitions, a consequence of the absence of a dark exciton state in MoSe$_2$/WSe$_2$ heterobilayers. The identification of the various neutral and charged exciton species provides new insight into multiple peaked IX spectra in heterobilayers, while the identification of the localized nature of IX ensemble emission in MoSe$_2$/WSe$_2$ heterobilayers demonstrate the important role of the moir{\'e} potential in their magneto-optical properties. These findings highlight the necessity to consider the spatial pinning of the IXs to the moir{\'e} lattice in order to achieve an accurate description of many-body exciton-exciton phenomena in TMD heterobilayers. 

    \section{Magneto-optics of spin-singlet and spin-triplet neutral IX\lowercase{s}}
    
\begin{figure*}
    	\begin{center}
    		\includegraphics[scale=0.6]{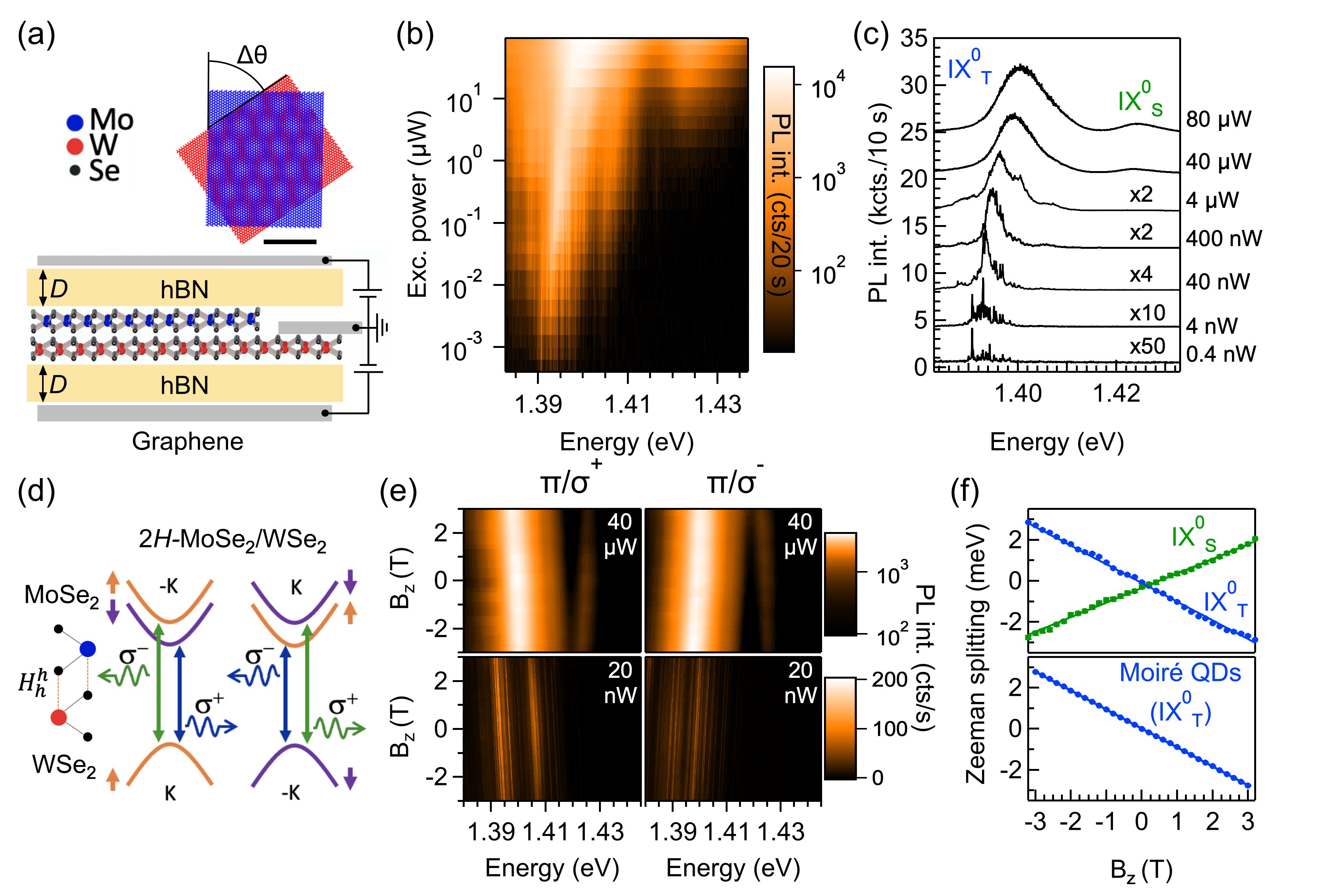}
    	\end{center}
        \caption{Power dependence of moir{\'e}-trapped IXs. \textbf{(a)} Sketch of the dual-gated WSe$_2$/MoSe$_2$ heterobilayer. Graphite layers are used as contacts for top, bottom, and heterobilayer, while hBN layers ($D=$ 18 nm) are used as dielectric spacers. \textbf{(b)} Color plot with the full evolution of the low-temperature (T = 4 K) confocal PL spectrum of a representative spot in the undoped heterobilayer for  0.4 nW $ \leq P_{exc} \leq$ 80 $\mu$W in logarithmic scale. \textbf{(c)} Linecuts extracted from the color plot in (b) for $P_{exc}$ with different orders of magnitude. The spectra have been shifted vertically and normalized by the values indicated in the figure for visualization purposes. \textbf{(d)} Schematics of the spin–valley configuration of the proposed optical transitions with the corresponding selection rules for $H_h^h$ atomic registry \cite{yu2018brightened}. \textbf{(e)} Magnetic-field dependence of IX PL in the low- and high-excitation regimes (bottom and top panels, respectively) for $\sigma^+$- and $\sigma^-$-resolved confocal collection (left and right panels, respectively). \textbf{(f)} Magnetic-field dependence of the Zeeman splitting measured for IX$^0_T$ and IX$^0_S$ at high $P_{exc}$ (top panel), and for a representative moir{\'e}-trapped IX at low $P_{exc}$.}
        \label{fig1}
    \end{figure*}

    Figure \ref{fig1}(a) shows a sketch of the dual-gated heterobilayer device we employ, which consists of a ML MoSe$_2$ and a ML WSe$_2$ vertically stacked with a twist angle $\Delta\theta\sim 56.4 \pm 0.2^\circ$ (2$H$ stacking). The twist angle in our heterobilayer is beyond the theoretically proposed critical angle for lattice reconstruction \cite{rosenberger2020twist,weston2020atomic,andersenexcitons}, ensuring minimal moir{\'e} domain formation. The heterobilayer was encapsulated by hexagonal boron nitride (hBN). Graphene layers act as electrical contacts for the top, bottom and heterobilayer gates (see Ref. \cite{baek2020highly} for more details). Via voltage-dependent reflectance spectroscopy, we observe clear signatures of strongly correlated electronic states in both the conduction (CB) and valence bands (VB) at numerous fractional filling values of the moir{\'e} superlattice (see Suppl. Note 1). Our observations, consistent with recent reports of correlated insulating states in angle aligned WSe$_2$/WS$_2$ heterobilayer samples \cite{tang2020simulation,regan2020mott,xu2020correlated,liu2020excitonic}, confirm the quality of our heterobilayer sample and the formation of a moir{\'e} superlattice. However, in-depth analysis of this experiment is beyond the scope of the current manuscript. Next, we investigate the evolution of the low-temperature (T = 4 K) confocal PL spectrum as a function of the IX density (with all gates grounded). The density of optically-generated IXs in our sample can be varied by changing the power of the continuous-wave excitation laser ($P_{exc}$) \cite{wang2019optical}. We excite resonantly to the 1$s$ state of the intralayer A exciton in ML MoSe$_2$ ($\lambda = 759$ nm), and scan $P_{exc}$ over 5 orders of magnitude. Figure \ref{fig1}(b) shows a color plot of the full evolution of the PL spectrum at a representative spot in the grounded heterobilayer for 0.4 nW $ \leq P_{exc} \leq$ 80 $\mu$W, while Fig. \ref{fig1}(c) presents linecuts extracted from Fig. \ref{fig1}(b) for $P_{exc}$ of different orders of magnitude. At low excitation powers ($P_{exc} \leq$ 10 nW), the PL spectrum reveals several discrete narrow emission lines with energies in the range ~1.39 - 1.41 eV, consistent with recently reported values for neutral IXs trapped in the moir{\'e} potential landscape \cite{seyler2019signatures,brotons2020spin}. Magneto-optical studies in WSe$_2$/MoSe$_2$ heterobilayers with 2$H$ stacking have shown that the moir{\'e}-trapped IXs arise from optical transitions involving the lowest spin-orbit-split CB of MoSe$_2$ at $\pm K$ \cite{seyler2019signatures,brotons2020spin, baek2020highly}. This observation leads to spin-triplet optical transitions for the trapped IXs (see Fig. \ref{fig1}(d)). Although such spin-flip transitions are normally forbidden in ML TMDs \cite{wang2018colloquium}, they can be brightened due to the selection rules dictated by the resulting interlayer atomic registry of the heterostructures, as theoretically predicted \cite{yu2018brightened} and experimentally shown \cite{ciarrocchi2019polarization,wang2019giant,joe2021electrically,zhang2019highly}. 
    
    For the lowest excitation powers employed, only a few tens of IXs are generated ($\sim$10 to 40 depending on the spatial position in the sample). Figure S1 shows the full evolution of the PL spectrum measured at a second spatial location of the heterobilayer in a similar $P_{exc}$ range. The PL measured at both spatial locations shows the same overall behavior under increasing excitation power. First, the emission intensity of the few discrete narrow lines saturates with increasing power (see Figs. S2(a) and S2(b)), hallmarks of few-level quantum-confined systems. Simultaneously, as $P_{exc}$ increases, more trapping sites are populated with IXs and we progressively lose the ability to resolve individual spectral lines, since they merge into an IX ensemble band. The IX ensemble band blue-shifts with increasing $P_{exc}$. At excitation powers of $\sim2 \mu$W, a second IX emission peak appears at higher energy and continuously blue-shifts with increasing $P_{exc}$. For the highest $P_{exc}$ used in our experiments (80 $\mu$W), the two IX peaks exhibit linewidths between 5 and 10 meV (depending on the spatial location on the sample) and consistently exhibit an energy splitting of $\sim$24 meV. 
    
    To unambiguously identify if these peaks arise from band-edge states at $\pm K$ and disentangle the spin–valley configuration of each IX emission band, we perform helicity-resolved magneto-optical spectroscopy measurements in the Faraday configuration under linearly polarized ($\pi$) excitation at 1.63 eV (759 nm). Figure \ref{fig1}(e) shows the magnetic-field ($B_z$) dependence of IX PL in the low- and high-excitation regimes (bottom and top panels, respectively) for $\sigma^+$- and $\sigma^-$-resolved confocal collection (left and right panels, respectively). A clear Zeeman splitting with increasing $B_z$ is observed for the IX PL in both excitation regimes. The low-energy ensemble band (IX$^0_T$) observed at high $P_{exc}$ shows the same polarization dependence with $B_z$ as the few individually resolved  moir{\'e}-trapped IXs. However, the high-energy ensemble band (IX$^0_S$) presents the opposite polarization. Figure \ref{fig1}(f) shows the $B_z$-dependence of the experimental Zeeman splitting ($\Delta E_z$) values measured for IX$^0_T$ and IX$^0_S$ (top panel) and a representative single moir{\'e}-trapped IX (bottom panel), as extracted from Lorentzian fits of the experimental data. The Zeeman splitting is defined as $\Delta E_z(B_z) = E^{\sigma^+}(B_z)-E^{\sigma^-}(B_z)= g\mu_BB_z$, with $g$ an effective $g$-factor, $\mu_B$ the Bohr magneton and $E^{\sigma^\pm}$ the $B_z$-dependent energies of the IX transitions with $\sigma^\pm$ polarization. Linear fits reveal $g$-factors of $-15.71\pm0.05$, $12.24\pm0.13$, and $-15.98\pm0.02$ for IX$^0_T$, IX$^0_S$ and the moir{\'e}-trapped single IX, respectively. Since the carrier spin, the valley index and the atomic orbital involved in the optical transitions are associated to a magnetic moment \cite{seyler2019signatures,nagler2017giant,xu2014spin,xiao2012coupled}, the measured $g$-factors provide valuable information about the nature of the transitions. In this regard, the band-picture model presented in Ref. \cite{aivazian2015magnetic} has successfully predicted both the magnitude and the sign of the IX $g$-factors in 2$H$- and 3$R$-stacked MoSe$_2$/WSe$_2$ heterobilayers \cite{seyler2019signatures, ciarrocchi2019polarization,wang2019giant,joe2021electrically,brotons2020spin,baek2020highly} that show excellent agreement with more rigorous theoretical descriptions of the IX Zeeman shifts \cite{wozniak2020exciton}. Taking into account the reported effective masses of $m^*_h = 0.37$ $m_0$ and $m^*_e = 0.84$ $m_0$ for holes in the topmost valence band of WSe$_2$ \cite{kormanyos2015k} and electrons in the lowest CB of MoSe$_2$ at at $\pm K$ \cite{larentis2018large,goryca2019revealing} (with $m_0$ the free electron rest mass), we estimate theoretical $g$-factor values of $-15.8$ and $11.8$ for IX$^0_T$ and IX$^0_S$ transitions, respectively \cite{brotons2020spin}. The good agreement between the experimental and calculated $g$-factor values confirms the spin-valley configurations of both excitonic transitions and corroborates the identification of IX$^0_T$ and IX$^0_S$ in the high excitation regime (see Fig. \ref{fig1}(d) for a level schematic, the optical transitions, and the corresponding selection rules \cite{yu2018brightened}), in agreement with recent works \cite{ciarrocchi2019polarization, wang2019giant,joe2021electrically,zhang2019highly}. These results bring to light a striking property of these IX transitions: unlike counterpart ML TMDs \cite{echeverry2016splitting,wang2018colloquium}, TMD heterobilayers do not host dark excitons (i.e. spin-forbidden optical transitions). In the rest of the work, we focus on the properties of the IX ensemble emission peaks with different spin configurations (IX$^0_T$ and IX$^0_S$) observed at high IX densities.

    \begin{figure*}
        \begin{center}
        \includegraphics[scale= 0.6]{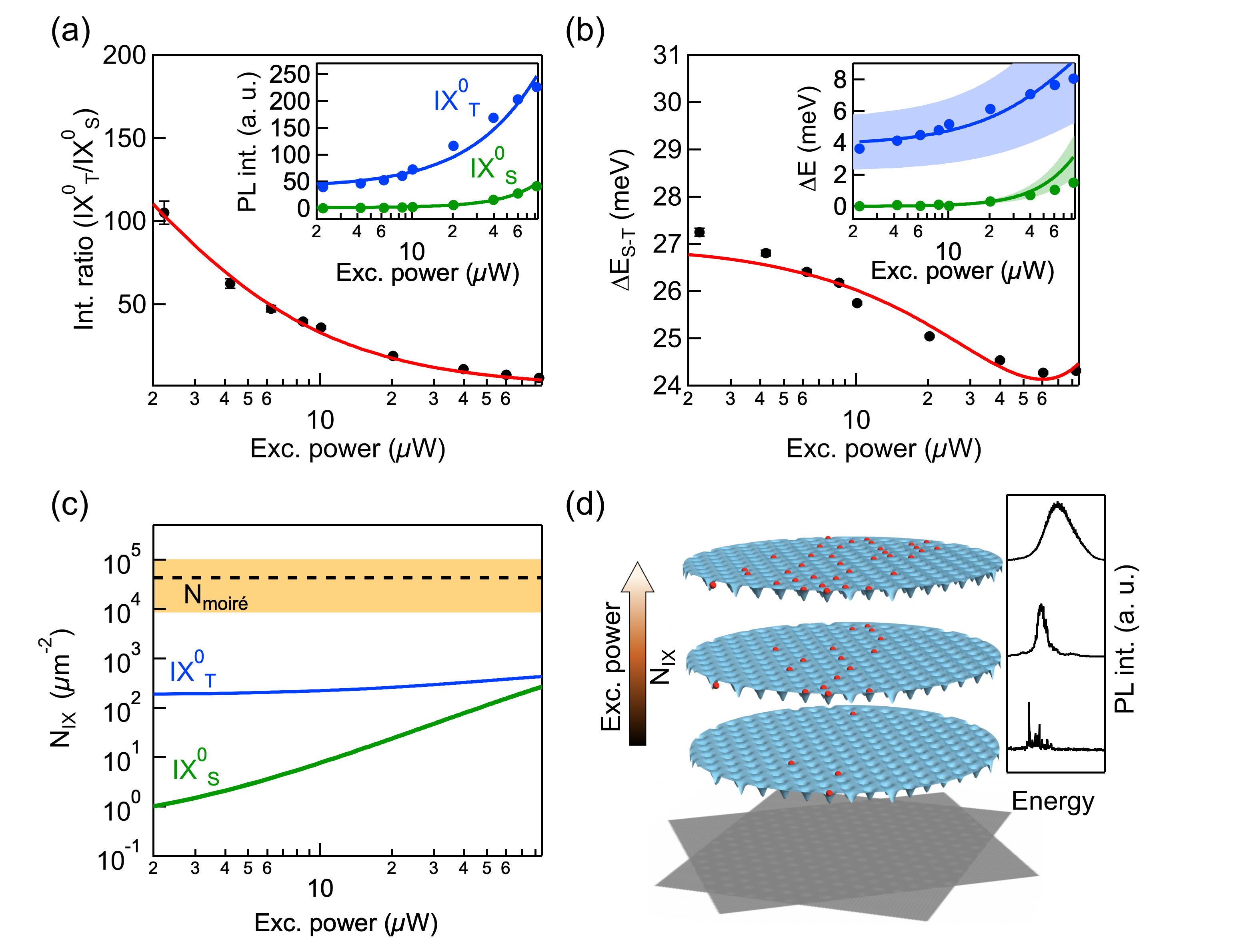}
        \end{center}
        \caption{Dipolar interactions of an ensemble of trapped IXs. \textbf{(a)} Ratio of integrated PL intensities between the IX$^0_T$ and IX$^0_S$ peaks (IX$^0_T$/IX$^0_S$) as a function of $P_{exc}$ extracted from Fig. \ref{fig1}(b) (black dots). The inset shows the evolution of the integrated PL intensities for each individual peak in the same $P_{exc}$ range. The red, blue and green solid lines represent fits of the experimental data to the rate equation model described in the Suppl. Mat. \textbf{(b)} Energy splitting between IX$^0_S$ and IX$^0_T$ ($\Delta E_{S-T}$) as a function of the excitation power (black dots). The inset shows the energy shifts of the individual peaks. The red solid line represents a fit of the experimental $\Delta E_{S-T}$ to Eq. (\ref{Eq:energy_splitting}), while the green and blue solid lines are fits of the measured $\Delta E_{S,T}$ to Eq. (\ref{Eq:plate_capacitor}) and the shaded areas represent the uncertainty interval corresponding to $d$ values ranging from 0.4 \cite{baek2020highly} to 1 nm \cite{nagler2017interlayer}. \textbf{(c)} Estimated power-dependent densities of IXs with spin-singlet (green) and spin-triplet (blue) configurations. The thickness of the lines indicates the confidence interval estimated from the fits to the experimental data. The black dashed line indicates the density of moir{\'e} traps estimated for our heterostructure, while the orange shaded area represents the density of sites for MoSe$_2$/WSe$_2$ heterobilayers with stacking angles between 54.4$^\circ$ and 58.4$^\circ$. \textbf{(d)} Sketch depicting the transition between the low- and high-excitation power regimes for IXs (red spheres) trapped in the moir{\'e} sites of a MoSe$_2$/WSe$_2$ heterobilayer. At low $P_{exc}$, only a few IXs are a localized in the moir{\'e} sites lying inside the circular confocal collection spot (not to scale). As $P_{exc}$ increases, more and more optically-generated IXs are trapped in the moir{\'e} sites, giving rise to a broad PL emission band originating from an ensemble of moir{\'e}-trapped IXs. Even at the highest $P_{exc}$ used in this work, less than $\sim2\%$ of moir{\'e} sites contain trapped IXs.}
      \label{fig2}
    \end{figure*}
    
    \section{Dipolar interactions of an ensemble of trapped IX\lowercase{s}}
    
    To understand the link between the individually resolved moir{\'e}-trapped IXs and the ensemble IX emission, we focus on the power-dependent blue-shifts observed in the PL energy of both IX$^0_T$ and IX$^0_S$. The peak blue-shifts originate from the repulsive dipolar exciton-exciton interaction of IXs, which arises as a consequence of the large permanent electrical dipole moment induced by the spatial separation of the exciton carriers \cite{butov1999magneto,nagler2017interlayer,wang2018electrical,unuchek2019valley}. The power-dependent energy shifts for IX$^0_T$ ($\Delta E_{T}$) and IX$^0_S$ ($\Delta E_{S}$) can be expressed as 
    $\Delta E_{S,T}(P_{exc}) = E_{S,T}(P_{exc})-E_{S,T}^0$, with $E_{S,T}(P_{exc})$ and $E_{S,T}^0$ being the energy of the corresponding exciton species at $P_{exc}$ and vanishing excitation, respectively. Such excitation-dependent energy shifts can be quantitatively estimated using the plate capacitor formula \cite{butov1999magneto,nagler2017interlayer}:
    \begin{align}
        \Delta E_{S,T}(P_{exc})=4\pi N_{S,T}(P_{exc})e^2d/\epsilon,
        \label{Eq:plate_capacitor}
    \end{align}
    where $N_{S,T}$ is the exciton density of the corresponding IX configuration, $e$ is the electron charge, $d$ is the interlayer distance, and $\epsilon$ is the dielectric constant (see Suppl. Note 3 for the derivation and applicability of Eq. (\ref{Eq:plate_capacitor})). Therefore, Eq. (\ref{Eq:plate_capacitor}) allows us to estimate the interlayer exciton density at different excitation powers from the experimentally measured energy blue-shift ($N_{S,T}=\Delta E_{S,T} \varepsilon/(4\pi e^2d)$) \cite{butov1999magneto,nagler2017interlayer,jauregui2019electrical,li2020dipolar}. For example, from the data shown in Fig. \ref{fig1}(b) we estimate a $\Delta E_{T}\sim0.56$ meV at $P_{exc}=4$ nW. Assuming an $\varepsilon = 7.4 \varepsilon_0$ \cite{gao2017interlayer} (with $\varepsilon_0$ the vacuum permittivity), and a $d$ ranging from $\sim0.4$ nm \cite{baek2020highly} to 1 nm \cite{nagler2017interlayer}, we estimate a $N_T(4$ nW$)\approx 1.8\cdot10^{9} - 4.6\cdot10^{9}$ cm$^{-2}$. The estimated value of $N_T$ suggests the presence of $\sim$18 - 46 trapped IX$_T^0$ in a spot of about 1 $\mu m^2$, which corresponds roughly to the area of the confocal spot in our measurements. The estimated number of trapped IX$_T^0$ agrees well with the magnitude of the total number of peaks that can be experimentally resolved at $P_{exc}=4$ nW. In a similar way, from the estimated total $\Delta E_{T}$ of $\sim$8.06 meV for IX$_T^0$ in Fig. \ref{fig1}(b), we estimate a $N_T\approx 2.6\cdot10^{10} - 6.6\cdot10^{10}$ cm$^{-2}$ at the highest excitation powers used in our experiments. For such IX densities, the IXs present an average exciton-exciton distance $\langle r_{IX}\rangle\sim 42-67$ nm (see Suppl. Note 4 for more details). The estimated $\langle r_{IX}\rangle$ in our experiments is of the same order of magnitude than the $\langle r_{IX}\rangle$ inferred in other 2D excitonic systems for which dipolar interactions are typically observed, for example indirect excitons in III-V quantum wells \cite{butov1999magneto} and MoSe$_2$/WSe$_2$ heterobilayers \cite{nagler2017interlayer,li2020dipolar,jauregui2019electrical}. 
    
   The estimated values of $N_T$ and $\langle r_{IX}\rangle$ shed some light on the nature of the IX PL emission bands observed at high $P_{exc}$. The stacking angle in our heterostructure ($\Delta\theta = 56.4 \pm 0.2^\circ$ \cite{baek2020highly}) gives rise to an estimated density $N_{total}\sim1.6\cdot10^{13}$ cm$^{-2}$ of moir{\'e} trapping sites with three different local atomic configurations: $H_h^h$, $H_h^X$ and $H_h^M$ \cite{yu2017moire}, where $H_{h}^{\mu}$ denotes an $H$-type stacking with either $h$ the hexagon centre, $X$ the chalcogen site or $M$ the metal site vertically aligned with the hexagon centre ($h$) of the hole layer. $N_{total}$ yields a density $N_{moir\Acute{e}}=N_{total}/3\sim4.2\cdot10^{12}$ cm$^{-2}$ of moir{\'e} sites with the same atomic registry. Therefore, our analysis reveals that the estimated density of optically-generated IXs ($N_{IX}$) is around two orders of magnitude smaller than $N_{moir\Acute{e}}$; less than 2$\%$ of the moir{\'e} sites are filled with IXs. We note that the highest $N_{IX}$ achieved in our experiments corresponds to the lowest $N_{IX}$ investigated by Wang \textit{et. al.}, who reported an optically driven Mott transition from IXs to a charge-separated electron and hole plasmas in a 3$R$-MoSe$_2$/WSe$_2$ heterobilayer with a similar moir{\'e} period ($\Delta\theta=4^{\circ}$) for $N_{IX}>$ $3\cdot10^{12}$ cm$^{-2}$ (i.e., for $N_{IX}>N_{moir\Acute{e}}$) \cite{wang2019optical}.
    
    In order to provide an additional degree of confidence in the IX density estimated from Eq. (\ref{Eq:plate_capacitor}), we estimate the magnitudes of $N_T$ and $N_S$ from the power dependent results shown in Fig. \ref{fig1}(b). Figure \ref{fig2}(a) shows the ratio of integrated PL intensities between the IX$^0_T$ and IX$^0_S$ peaks as a function of $P_{exc}$ extracted from Fig. \ref{fig1}(b) (black dots). The inset shows the evolution of the integrated PL intensities for each individual peak in the same $P_{exc}$ range. In the range of $P_{exc}$ for which both peaks coexist, we observe that the intensity ratio decreases from $\sim$105 to $\sim$5 with increasing $P_{exc}$. To evaluate the PL intensities of the two IX species peaks quantitatively, we employ a rate equation model (see Suppl. note 5). The solid lines in Fig. \ref{fig2}(a) represent fits of the experimental data for both the PL intensity ratio (red) and the PL intensities of the IX$^0_T$ (blue) and IX$^0_S$ (green) peaks to the rate equation model. The fits allow us to confidently estimate the power-dependent relative densities of IXs with spin-singlet and spin-triplet configurations. Next, we show in Fig. \ref{fig2}(b) the experimentally-measured energy splitting between IX$^0_S$ and IX$^0_T$ ($\Delta E_{S-T}$), where we observe that increasing $P_{exc}$ gives rise to a reduction of $\Delta E_{S-T}$ by up to 3 meV, well beyond the uncertainty associated to the experimental determination of the energy splitting. From Eq. (\ref{Eq:plate_capacitor}), the density-dependent $\Delta E_{S-T}$ can be calculated as
    \begin{align}
        \Delta E_{S-T}=\Delta E_{S-T}^0 +\Delta N_{S-T}(P_{exc})4\pi e^2d/ \varepsilon,
        \label{Eq:energy_splitting}
    \end{align}
    with $\Delta E_{S-T}^0=E_S^0-E_T^0$ and $\Delta N_{S-T}(P_{exc})=N_{S}(P_{exc})-N_{T}(P_{exc})$. From Eq. (\ref{Eq:energy_splitting}) it is straightforward to see that the decrease of $\Delta E_{S-T}$ with increasing $P_{exc}$ has its origin in the higher density of IX$^0_T$ in the range of $P_{exc}$ used in our experiments, which leads to $\Delta N_{S-T}(P_{exc})<0$. Figure \ref{fig2}(b) shows the fit of the experimental $\Delta E_{S-T}$ (black dots) to Eq. (\ref{Eq:energy_splitting}) (red solid line), where we have used the relative IX densities shown in \ref{fig2}(c), and have assumed an average $d$ value of 0.7 nm \cite{li2020dipolar}. The inset shows a comparison of the experimental and calculated energy shift for each individual exciton band, where the shaded areas represent the confidence interval corresponding to energy shifts calculated using $d$ values ranging from 0.4 \cite{baek2020highly} to 1 nm \cite{nagler2017interlayer}. The good agreement between the experimental and calculated values allows us to estimate the order of magnitude of the absolute densities for IXs with both spin configurations, which are shown in Fig. \ref{fig2}(c).
    
    \begin{figure*}[t]
    	\begin{center}
    		\includegraphics[scale=0.6]{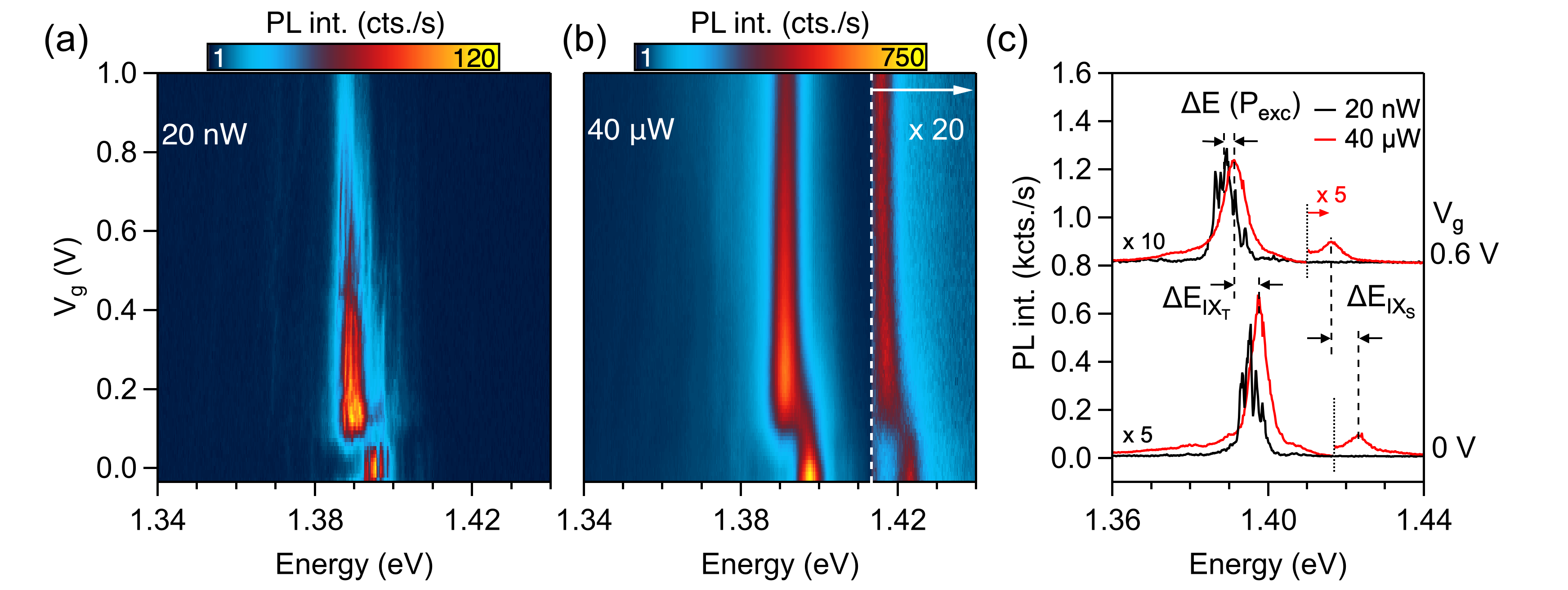}
    	\end{center}
        \caption{Interlayer exciton trions in WSe$_2$/MoSe$_2$. \textbf{(a), (b),} Gate-voltage controlled PL of IXs in the neutral ($0< V_g \leq0.1$ V) and electron doping regime ($V_g \geq 0.1$ V) at $B_z$ = 0 T, for $P_{exc}=20$ nW (a) and $P_{exc}=40$ $\mu$W (b) at 4 K. In (b), the PL intensity has been multiplied by a factor 20 in the spectral range delimited by the vertical dashed area for visualization purposes. \textbf{(c)}, Linecuts extracted from the color plots in (a) and (b) (black and red solid lines, respectively) in the neutral ($V_g=0$ V) and electron doping regimes ($V_g=0.6$ V). The spectra have been normalized by the values indicated in the figure, and the linecuts at $V_g=0.6$ V have been shifted vertically for visualization purposes. The dashed vertical lines indicate the central energies of the different ensemble peaks, illustrating the power-induced blue-shift ($\Delta E(P_{exc})$) and the energy splittings arising from the binding energies of the charged IXs with spin triplet ($\Delta E_{IX_T}$) and spin singlet ($\Delta E_{IX_S}$) configuration.}
        \label{fig3}
    \end{figure*}
    
    The black dashed line in Fig. \ref{fig2}(c) shows the estimated $N_{moir\Acute{e}}$ for our sample. In agreement with the analysis from the plate-capacitor approximation (Eq. (\ref{Eq:plate_capacitor})), the results in Fig. \ref{fig2}(c) corroborate that, even for the highest excitation powers used in our experiments, the estimated densities of optically-generated IXs are around two orders of magnitude smaller than $N_{moir\Acute{e}}$. The orange shaded area in Fig. \ref{fig2}(c) represents the estimated $N_{moir\Acute{e}}$ for MoSe$_2$/WSe$_2$ heterobilayers with stacking angles between 54.4$^\circ$ and 58.4$^\circ$, showing that this conclusion holds true even for a large range of stacking angles. 
    
    Our results represent a bridge between the quantum-dot-like \cite{seyler2019signatures,brotons2020spin,baek2020highly} and the broad PL emission peaks \cite{hanbicki2018double,ciarrocchi2019polarization,wang2019giant,jauregui2019electrical,miller2017long,sigl2020signatures,choi2021twist} previously observed for IX ensembles in MoSe$_2$/WSe$_2$ heterobilayers at different excitation powers. Figure \ref{fig2}(d) shows a sketch depicting the transition between the low- and high-density regimes. At low $P_{exc}$, only a few IXs (red spheres) are localized in the moir{\'e} sites lying inside our circular confocal collection spot (not to scale). The spatial and spectral isolation of these trapped excitons is likely aided by dipolar repulsion, which minimizes the probability for excitons to populate neighboring moir{\'e} sites. As $P_{exc}$ increases, more and more optically-generated IXs are trapped in the moir{\'e} sites, giving rise to a broad PL emission band originating from the ensemble of moir{\'e}-trapped IXs. Even at the highest $P_{exc}$ used in this work, less than 2$\%$ of moir{\'e} sites contain trapped IXs. This behaviour agrees well with the magneto-optical properties measured for single confined IXs and the IX$^0_T$ ensemble PL band. The polarization selection rules of moir{\'e}-trapped excitons are dictated by the local atomic registry of the moiré trapping site \cite{yu2017moire,yu2018brightened}. The same optical selection rules observed for both single confined IXs and the IX$^0_T$ ensemble PL band indicate that only moir{\'e} sites with a local atomic registry $H_h^h$ are responsible for the IX trapping \cite{seyler2019signatures,zhang2019highly,brotons2020spin,baek2020highly}. Moreover, the optical selection rules of the IX$^0_S$ ensemble PL peaks further corroborate this affirmation, since $H_h^h$ is the only local atomic registry in $2H$-stacked TMD heterobilayers that results in opposite circularly-polarized transitions for IXs with spin-triplet and spin-singlet configurations \cite{yu2018brightened}. We note that for MoSe$_2$/WSe$_2$ heterobilayers with $3R$-stacking (i.e. $\Delta\theta\sim0^\circ$), the magneto-optical selection rules suggest that the PL emission at high excitation power arises from spin-singlet and spin-triplet IXs with a local stacking registry $R^X_h$ \cite{ciarrocchi2019polarization, joe2021electrically}.
    
    \section{{Spin-valley configurations of negative IX trions in 2$H$-M\lowercase{o}S\lowercase{e}$_2$/WS\lowercase{e}$_2$}}
    
    \begin{figure*}[t]
    	\begin{center}
    		\includegraphics[scale=0.6]{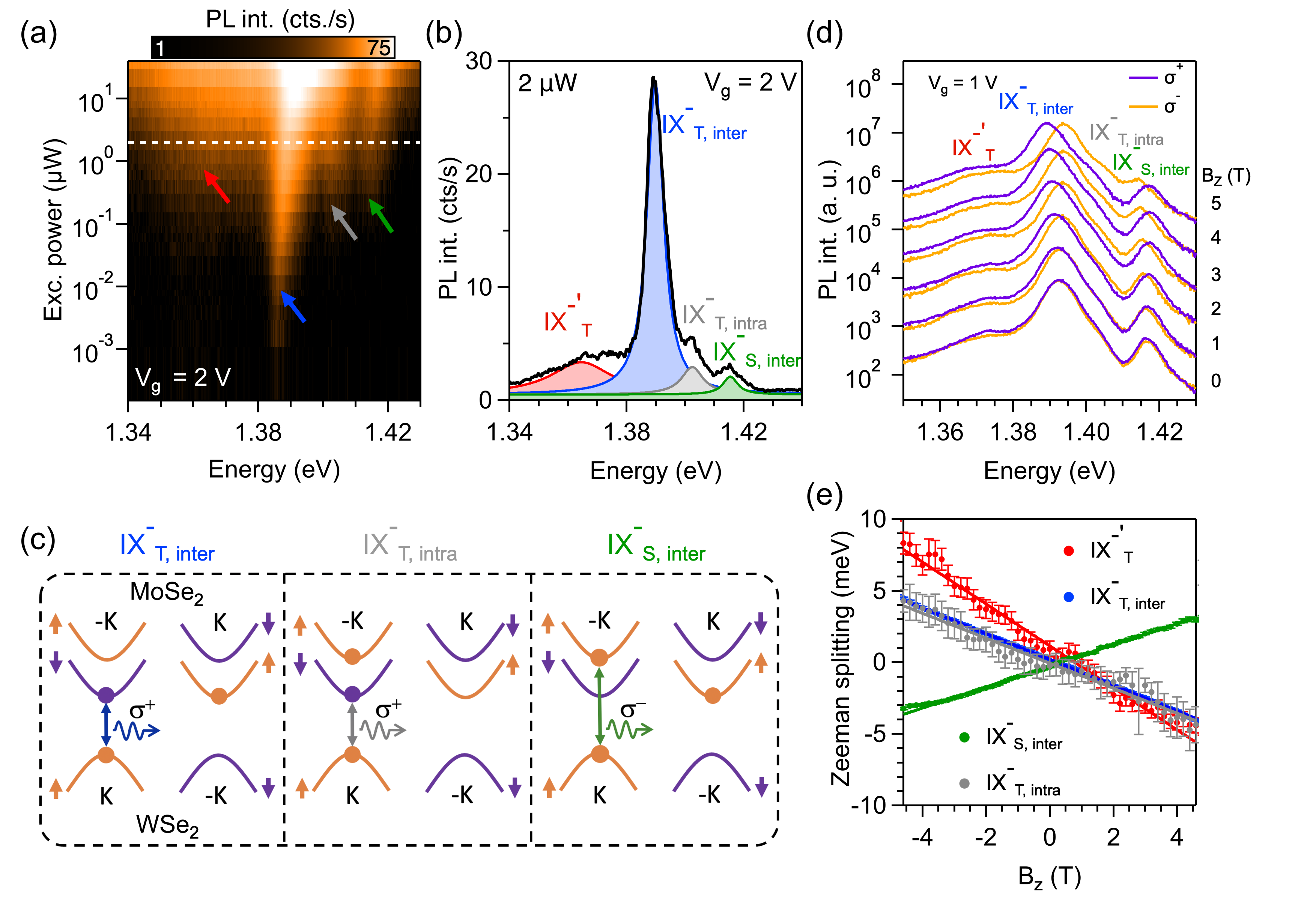}
    	\end{center}
        \caption{Interlayer exciton trions in WSe$_2$/MoSe$_2$. \textbf{(a)} Color plot with the full evolution of the PL spectrum of negative IX trions for  0.2 nW $ \leq P_{exc} \leq$ 40 $\mu$W at $V_g=2$ V in logarithmic scale, in which four different peaks can be resolved (as indicated by the arrows). \textbf{(b)} PL spectrum acquired for the intermediate $P_{exc}$ value indicated by the white dashed line in (a) (2 $\mu$W) at $V_g=2$ V, fitted to four lorentzian peaks corresponding to various exciton species: IX$^-_{T, inter}$ (blue), IX$^-_{T,intra}$ (gray), IX$^-_{S,inter}$ (green), and IX$^{-'}_T$ (red). \textbf{(c)} Schematic representation of the charge configurations for IX$^-_{T, inter}$ (left), IX$^-_{T,intra}$ (middle) and IX$^-_{S,inter}$ (right) showing the optical transitions that involve a hole in the the topmost valence band of WSe$_2$ at $K$. \textbf{(d)} Magnetic field dependence of the IX trions PL for $\sigma^+$- (purple) and $\sigma^-$-polarized (orange) collection in the range $0\leq B_z\leq5$ T under $V_g=$ 1 V and $P_{exc}=40$ $\mu$W. \textbf{(e)} Zeeman splitting of each IX exciton species (dots) as extracted from fits of the experimental data shown in (d). The solid lines represent linear fits of the experimental data.}
        \label{fig4}
    \end{figure*}
    
    In this section, we investigate the formation of negatively charged trapped IXs with both spin-triplet and spin-singlet configuration. The Fermi energy in our WSe$_2$/MoSe$_2$ heterobilayer can be continuously tuned with an external gate voltage $V_g$. The color plots of Figs. \ref{fig3}(a) and \ref{fig3}(b) show the effects of electron doping on the PL of the trapped IXs at low ($P_{exc}=20$ nW) and high ($P_{exc}=40\mu$W) excitation powers, respectively. The $V_g$-dependent evolution of the IX PL shows the same overall behaviour for both low and high IX density regimes, and resembles the doping dependence of the intralayer A exciton of monolayer MoSe$_2$ measured in a sample region where the MoSe$_2$ does not overlap with the WSe$_2$ layer (see Suppl. Note 6). For $0< V_g \lesssim 0.1$ V, the PL spectrum is dominated by neutral excitons: IX$^0_T$ at low IX density (Fig. \ref{fig3}(a)), and both IX$^0_T$ and IX$^0_S$ at high IX density (Fig. \ref{fig3}(b)). At high IX density, the IX$^0_T$ and IX$^0_S$ ensemble peaks present linewidths of $\sim5.2$ meV and $\sim7$ meV, respectively, on par with the cleanest MoSe$_2$/WSe$_2$ samples \cite{ciarrocchi2019polarization,calman2020indirect,sigl2020signatures}. For $V_g \gtrsim 0.1$ V, new red-shifted peaks appear at energies $\sim$7 ($\sim$4) meV below the IX$^0_T$ (IX$^0_S$) peaks, indicating the formation of negatively charged IX trions with different spin configurations (see linecuts in Fig. \ref{fig3}(c)). The measured energy differences between the neutral and charged exciton peaks are attributed to the binding energies of the charged IXs, in good agreement with recently reported values for MoSe$_2$/WSe$_2$ heterostructures with $3R$ stacking \cite{jauregui2019electrical,joe2021electrically}.  The new trion peaks coexist with IX$^0_T$ and IX$^0_S$ for a short range of applied voltages until IX$^0_T$ and IX$^0_S$ eventually vanish with increasing $V_g$. We note the observation of IX trions in the quantum emitter (low-excitation) regime is novel. At high excitation powers, the PL spectrum in the $n$-doped region shows broad emission shoulders in both the low- and high-energy tails of the IX trion peak with spin-triplet configuration. In order to spectrally resolve the emission bands, we plot the PL of IXs for $n$ doping ($V_g=2$ V) as a function of $P_{exc}$ (see Fig. \ref{fig4}(a)). Figure \ref{fig4}(b) shows the PL spectrum acquired for an intermediate $P_{exc}$ value of 2 $\mu$W at $V_g=2$ V, as indicated by the white dashed line in Fig. \ref{fig4}(a). At this excitation power, we clearly resolve four emission peaks. The brightest peak in the spectrum (at $\sim$ 1.39 eV) corresponds to intervalley IX trions with spin-triplet optical transitions (IX$^-_{T, inter}$). At higher emission energies, we observe two additional peaks with an energy splitting of $\sim$12 meV. These two peaks show a similar integrated PL intensity across the entire range of $P_{exc}$ (see Suppl. Note 7), suggesting that their corresponding charge configurations involve the same band-edge energy levels. We attribute the low energy and high energy peaks of this doublet to intravalley (IX$^-_{T,intra}$) and intervalley (IX$^-_{S,inter}$) trions with spin-triplet and spin-singlet configuration, respectively. Figure \ref{fig4}(c) shows a schematic representation of the charge configurations for IX$^-_{T, inter}$ (left), IX$^-_{T,intra}$ (middle), and IX$^-_{S,inter}$ (right) with optical transitions that involve a hole in the the topmost valence band of WSe$_2$ at $K$. 
    
    The PL spectrum in Fig. \ref{fig4}(b) also shows a broad emission peak at lower energies than IX$^-_{T, inter}$ (IX$^{-'}_T$). Although its origin is not yet fully understood, this feature has also recently observed in the reflectivity of $n$-doped ML WSe$_2$ \cite{van2019probing, wang2020observation} and the PL of $n$-doped MoS$_2$ \cite{roch2020first,klein2021controlling} and attributed to  either a Mahan-like exciton \cite{roch2020first} or an exciton-plasmon-like excitation \cite{van2019probing, wang2020observation}. We find that contrary to the behaviour of IX$^-_{T, inter}$, IX$^-_{S,inter}$ and IX$^-_{T,intra}$, the integrated emission intensity of IX$^{-'}_T$ increases with the carrier concentration. Moreover, increasing the electron concentration also leads to a linear increase of the energy splitting between IX$^{-}_{T,inter}$ and IX$^{-'}_T$ (see Suppl. Note 8).
    
    Finally, the charge configurations for the different charged exciton species depicted in Fig. \ref{fig4}(c) highlight a striking difference between IX$^-_{S,inter}$ and 
    IX$^-_{T,intra}$. Unlike IX$^-_{S,inter}$, in which the photon emission arises from electron-hole recombination between the topmost VB and the top spin-split CB, the absence of a dark exciton ground state in MoSe$_2$/WSe$_2$ heterostructures enables the electron-hole recombination between the topmost VB and the bottommost spin-split CB for IX$^-_{T,intra}$. This behaviour contrasts with negative trions in W-based TMDs, in which both intravalley and intervalley negative trions present electron-hole recombination involving the same top spin-plit CB \cite{lyons2019valley}. Moreover, since IX$^-_{S,inter}$ is the only trion species in which the electron-hole recombination involves the top spin-split CB, the $V_g$-dependent relative integrated PL intensities of the different trion species should provide an estimate of the energy splitting between the spin-split CBs in MoSe$_2$  \cite{joe2021electrically}. We observe that the PL intensity from IX$^-_{S,inter}$ overtakes the combined PL intensity from IX$^-_{T, inter}$ and IX$^-_{T,intra}$ at $V_g \sim$ 6.1 V (see Suppl. Note 9), which corresponds to an energy splitting of $\sim$ 21 meV, in good agreement with the calculated spin-orbit splitting of MoSe$_2$ CBs at $\pm K$ (23 meV) \cite{kormanyos2014spin}. Therefore, the spin-valley configurations depicted in Fig. \ref{fig4}(c) for the different trion states suggest that, contrary to intravalley negative trions in W-based TMDs, although IX$^-_{S,inter}$ and IX$^-_{T,intra}$ involve the topmost CB, they should present different magneto-optical properties.
    
    To investigate the magneto-optical properties of the different trion peaks, we measure their PL as a function of $B_z$ at $P_{exc}=40$ $\mu$W and $V_g=$ 1 V, for which both the intensity and linewidth of the different peaks allow us to clearly monitor their emission properties. Figure \ref{fig4}(d) shows the magnetic field dependence of the IX trions PL for $\sigma^+$- (purple) and $\sigma^-$-polarized (orange) collection in the range $0\leq B_z\leq5$ T. The results reveal a clear Zeeman splitting for the three IX trion species: a positive vertical $B_z$ field leads to a red-shift of the $\sigma^+$-polarized emission for both IX$^-_{T, inter}$ and IX$^{-}_{T,intra}$, while it induces a slight blue-shift of the $\sigma^+$-polarized emission for IX$^-_{S,inter}$. Such contrasting behaviour can also be seen in the experimental Zeeman splitting of each IX trion species as shown in Fig. \ref{fig4}(e): IX$^-_{T, inter}$ and IX$^-_{T,intra}$ present negative slopes (negative $g$-factor), whereas IX$^-_{S,inter}$ exhibits a positive one. Linear fits of the measured Zeeman splittings reveal $g$-factors of $-15.44\pm0.06$, $-15.1\pm0.4$ and $12.31\pm0.11$ for IX$^-_{T, inter}$, IX$^-_{T,intra}$ and IX$^-_{S,inter}$, respectively. The extracted values show that IX$^-_{T, inter}$ and IX$^-_{S,inter}$ have $g$-factors with opposite signs and different magnitudes, as previously observed for IX$^0_T$ and IX$^0_S$ in the results shown in Fig. \ref{fig1}(f). More importantly, the fits also reveal that IX$^-_{T, inter}$ and IX$^-_{T,intra}$ present $g$-factors with the same sign and very similar magnitude, which confirms that the electron-hole recombination in these trions involve the same exact CB and VB, confirming the origin of the IX$^-_{T,intra}$ peak. We note that the particular band alignment and optical selection rules in 2$H$-MoSe$_2$/WSe$_2$ heterobilayers allow the formation of intervalley and intravalley IX trions by optical pumping of both the upper and lower CBs of MoSe$_2$ even in the neutral doping regime (see Suppl. Note 10). These results suggest the importance of revisiting the interpretation of multiple IX emission peaks in previous reports. Finally, the fits reveal a dependence of the trions $g$-factors with the carrier concentration. Figure S7 shows the evolution of the $g$-factor of the brightest peak in our spectra (IX$^-_{T, inter}$) as a function of electron doping. Similar to the case of the $g-$factor of quasiparticles in a 2D electron gas \cite{janak1969g}, and intralayer exciton polarons in ML TMDs \cite{roch2019spin, wang2018strongly, back2017giant}, we observe a change of $g$ with increasing electron doping. This change in the effective $g$-factors with increasing carrier doping has been attributed to many-body interactions and phase space filling effects \cite{back2017giant,wang2018strongly}. 
    
    Figure \ref{fig4}(e) also shows the measured Zeeman splitting for the excitonic feature IX$^{-'}_T$. The observed negative slope in the Zeeman splitting of this peak with increasing $B_z$ reveals similar selection rules to IX$^-_{T, inter}$. However, similar to the exact origin of IX$^{-'}_T$, the extracted $g$-factor of $-25.3\pm0.5$ is not yet  understood. Finally, we note a quadratic red-shift of the central energy of the Zeeman-split peaks of the IX ensemble peaks with increasing magnetic field (see Suppl. Note 12), which is absent for low IX densities, highlighting the importance of the strong exciton-exciton interactions at high IX densities.
     
    \section{Conclusion and Perspectives}
    
    In summary, we investigate neutral and negatively charged IXs trapped in moir{\'e} confinement potentials in a gate-tunable 2$H$-WSe$_2$/MoSe$_2$ heterostructure. The homogeneous linewidths of the ensemble IX peaks we find in our sample ($\sim$ 5 meV for neutral and charged species) are on par with the best reported, confirming the high-quality nature of the sample thanks to hBN encapsulation and control of the Fermi-level. By scanning the excitation power over five orders of magnitude, we tune the density of optically generated neutral excitons and observe a continuous evolution of the trapped IX density, from a few isolated quantum emitters to a large ensemble. In the high-excitation regime, neutral IXs with spin-triplet (IX$^0_T$) and spin-singlet (IX$^0_S$) configurations are identified.  The IX ensemble peaks energetically blue-shift with increasing density due to dipolar repulsion. From this effect, we are able to estimate $N_{IX}$ and find that even at the highest IX density we measure, $N_{IX} << N_{moir\Acute{e}}$. We discover that the magneto-optical properties of the trapped IXs are identical, regardless of their density: both the narrow quantum-dot-like and broad ensemble IX emission originate from IXs localised in moir{\'e} potential traps with the same local atomic registry ($H_h^h$). Moreover, the optical selection rules of the IX$^0_S$ ensemble further corroborate this affirmation, since $H_h^h$ is the only local atomic registry in $2H$-stacked TMD heterobilayers that results in opposite circularly-polarized transitions for IXs with spin-triplet and spin-singlet configurations. 
    
    Next, the Fermi energy was tuned to create negatively charged moir{\'e} trapped IXs, which we observe in both the low- and high-IX-density regimes. To our knowledge, the observation of charged moir{\'e} trapped IXs in the low-density regime is novel, and an exciting avenue for future investigations. Binding energies for the on-site negatively charged IX are found to be 7 meV (4 meV) for the spin-triplet  (spin-singlet) trion configuration. Our magneto-optical measurements at high IX densities reveal a fine structure for negative IX trions with spin-triplet configuration; we clearly resolve intravalley (IX$^-_{T,intra}$) and intervalley (IX$^-_{T,inter}$) IX trions. We note that, using moderate excitation powers, the formation of intervalley and intravalley IX trions is observed even in the neutral-doping regime, which creates a multi-peaked PL spectrum consisting of six possible  neutral and charged IX species (IX$^0_T$, IX$^0_S$, IX$^-_{T,intra}$, IX$^-_{T,inter}$, IX$^-_{S,intra}$, and IX$^{-'}_T$). These results suggest it could be fruitful to revisit previous interpretations of multiple peaked IX spectra from MoSe$_2$/WSe$_2$ heterobilayers in previous reports, particularly those using ungrounded devices and high excitation powers.  
    
    We remark that the unified picture of narrow linewidth IX emitters and broad ensemble IX peaks reconciles contrasting IX spectra reported from nominally similar WSe$_2$/MoSe$_2$ heterobilayer samples. This unified picture provides further valuable evidence about the nature of quantum emitters in moir{\'e} heterostructures, complementing previous results \cite{seyler2019signatures,brotons2020spin,baek2020highly}. So, although the narrow linewidth peaks have an inhomogeneous energy distribution similar to defect-related single photon emitters in 2D materials, the properties of the moir{\'e} quantum emitters are identical to the ensemble IX and arise from the intrinsic symmetry of the moir{\'e} confining potential at the specific atomic registry. Further, in the low-excitation regime, spatial isolation of trapped IXs is likely aided by dipolar repulsion, which minimizes the probability for excitons to populate neighboring moir{\'e} sites. Finally, in the high-excitation regime, the results highlight the importance of considering the spatial pinning of the ensemble IXs to the moir{\'e} lattice in order to achieve an accurate description of many-body exciton-exciton phenomena in TMD heterobilayers: the dipole interactions will depend on, and can be tuned by, the twist angle.
    
    
    %

    \begin{acknowledgments}
        We thank Mikhail M. Glazov for fruitful discussions. This work is supported by the EPSRC (grant no. EP/P029892/1 and EP/L015110/1), the ERC (grant no.  725920) and the EU Horizon 2020 research and innovation program under grant agreement no. 820423. Growth of hBN crystals by K.W. and T.T. was supported by the Elemental Strategy Initiative conducted by the MEXT, Japan, Grant No. JPMXP0112101001, JSPS KAKENHI Grant No. JP20H00354, and the CREST (JPMJCR15F3), JST. B.D.G. is supported by a Wolfson Merit Award from the Royal Society and a Chair in Emerging Technology from the Royal Academy of Engineering.
        
        \textit{Note added in proof}.—Recently, two related works on the topic of moir{\'e}-trapped interlayer trions have become available \cite{liu2021signatures,baek2020coulomb}.
    \end{acknowledgments}

\end{document}